\documentclass[aps,prl,twocolumn,showpacs,superscriptaddress]{revtex4}
\usepackage{pstricks,pst-node,pst-text,pst-3d,pst-plot}

\usepackage{graphicx}
\usepackage[english]{babel}
\usepackage{bm}    %% For bold math symbols
\usepackage{eepic} %% For defining plot symbols \fancyplus etc

\def\reff#1{(\ref{#1})}
\newcommand{\be}{\begin{equation}}
\newcommand{\ee}{\end{equation}}
\newcommand{\<}{\langle}
\renewcommand{\>}{\rangle}

\def\spose#1{\hbox to 0pt{#1\hss}}
\def\ltapprox{\mathrel{\spose{\lower 3pt\hbox{$\mathchar"218$}}
 \raise 2.0pt\hbox{$\mathchar"13C$}}}
\def\gtapprox{\mathrel{\spose{\lower 3pt\hbox{$\mathchar"218$}}
 \raise 2.0pt\hbox{$\mathchar"13E$}}}

\newcommand{\scre}{{\cal E}}

\newcommand{\scrl}{{\cal L}}
\newcommand{\scrm}{{\cal M}}
\newcommand{\scrms}{{\cal M}_{\rm stagg}}

\newcommand{\scro}{{\cal O}}

\newcommand{\scrt}{{\cal T}}

\newsavebox{\fancyplusb}

\savebox{\fancyplusb}(2,2){
\thinlines
\put(0,-1){\line(0,1){2}}
\put(-1,-0.5){\line(0,1){1}}
\put(-1,0){\line(1,0){2}}
\put(1,-0.5){\line(0,1){1}}
\put(-0.5,-1){\line(1,0){1}}
\put(-0.5,1){\line(1,0){1}}
}

\newcommand{\fancyplus}{\begin{picture}(2,2)
\usebox{\fancyplusb}
\end{picture}}

\begin{document}

\title{%% Entropically-driven phase transition in the \\
       Phase transition in the
       3-state Potts antiferromagnet on the diced lattice}

\author{Roman Koteck\'y}
%%\email{kotecky@cucc.ruk.cuni.cz}
\affiliation{Center for Theoretical Study, Charles University, Prague, CZECH REPUBLIC}
\affiliation{Mathematics Institute, University of Warwick, Coventry CV4 7AL, UK}
\author{Jes\'us Salas}
%%\email{jsalas@math.uc3m.es}
% \affiliation{Instituto Gregorio Mill\'an and
%   Grupo de Modelizaci\'on, Simulaci\'on Num\'erica y Matem\'atica Industrial,
%   Universidad Carlos III de Madrid,
%   Avda.\ de la Universidad 30,
%   28911 Legan\'es, SPAIN}
\affiliation{
  %% Grupo de Modelizaci\'on, Simulaci\'on Num\'erica y Matem\'atica Industrial,
  %%  M.S.M.I.~Group,
  Gregorio Mill\'an Institute,
  Universidad Carlos III de Madrid,
  28911 Legan\'es, SPAIN}
\author{Alan D. Sokal}
%%\email{sokal@nyu.edu}
\affiliation{Department of Physics, New York University,
      4 Washington Place, New York, NY 10003, USA}
\affiliation{Department of Mathematics,
      University College London, London WC1E 6BT, UK}

%%\date{\today {\bf\  --- Need to fix final date!!!}}
\date{February 14, 2008; revised April 22, 2008}

\begin{abstract}
% We prove that
% %%, contrary to theoretical expectations,
% the 3-state Potts antiferromagnet on the diced lattice
% (dual of the Kagom\'e lattice)
% has an entropically-driven phase transition at nonzero temperature.
% We then present Monte Carlo simulations,
% using a cluster algorithm,
% of the 3-state and 4-state models.
% The 3-state model has a phase transition at
% $v = e^J - 1 = -0.860599 \pm 0.000004$
% %% ($J = -1.97040 \pm 0.00003$)
% that appears to be in the universality class of the 3-state Potts ferromagnet.
% %% We estimate the critical exponents $\nu = ?.??? \pm 0.???$,
% %% $\gamma_{\rm stagg}/\nu = ???? \pm ????$ and
% %% $\gamma_{\rm u}/\nu = ???? \pm ????$.
% The 4-state model is disordered throughout the physical region,
% including at zero temperature.
%
We prove that the 3-state Potts antiferromagnet on the diced lattice
(dual of the Kagom\'e lattice)
has entropically-driven long-range order at low temperatures
(including zero).
We then present Monte Carlo simulations, using a cluster algorithm,
of the 3-state and 4-state models.
The 3-state model has a phase transition to the high-temperature
disordered phase at $v = e^J - 1 = -0.860599 \pm 0.000004$
that appears to be in the universality class of the 3-state Potts ferromagnet.
The 4-state model is disordered throughout the physical region,
including at zero temperature.
\end{abstract}



\maketitle

The $q$-state Potts model \cite{Potts_52,Wu_82+84}
plays an important role in the theory of critical phenomena,
especially in two dimensions \cite{Baxter_book,Nienhuis_84,DiFrancesco_97},
and has applications to various condensed-matter systems \cite{Wu_82+84}.
Ferromagnetic Potts models are by now fairly well understood,
thanks to universality;
% and much is known about their phase diagrams \cite{Wu_82,Wu_84}
% and critical exponents \cite{Itzykson_collection,DiFrancesco_97,Nienhuis_84}.
but the behavior of antiferromagnetic Potts models
depends strongly on the microscopic lattice structure,
so that many basic questions must be investigated case-by-case:
Is there a phase transition at finite temperature, and if so, of what order?
What is the nature of the low-temperature phase(s)?
If there is a critical point, what are the critical exponents and the
universality classes?
Can these exponents be understood (for two-dimensional models)
in terms of conformal field theory \cite{DiFrancesco_97}?

One expects that for each lattice ${\cal L}$ there
exists a value $q_c({\cal L})$ such that
for $q > q_c({\cal L})$  the model has exponential decay
of correlations uniformly at all temperatures, including zero temperature,
while for $q = q_c({\cal L})$ the model has a zero-temperature critical point.
For $q < q_c({\cal L})$  any behavior is possible;
often (though not always) the model has a phase transition
at nonzero temperature, which may be of either first or second order
\cite{footnote_qc_exceptions}.
The first task, for any lattice, is thus to determine $q_c$.

Some two-dimensional antiferromagnetic models at zero temperature
have the remarkable property that they can be mapped onto a ``height''
(or ``interface'' or ``SOS-type'') model \cite{Salas_98}.
Experience tells us that when such a representation exists,
the corresponding zero-temperature spin model is most often critical
\cite{height_rep_exceptions}.
The long-distance behavior is then that of a massless Gaussian
with some ({\em a priori}\/ unknown) ``stiffness'' $K > 0$.
The critical operators can be identified via the height mapping,
and the corresponding critical exponents can be predicted in terms
of the single parameter $K$.
Height representations thus provide a means for recovering
a sort of universality for some (but not all) antiferromagnetic models
and for understanding their critical behavior
in terms of conformal field theory \cite{DiFrancesco_97}.
% All the nonuniversal details of the microscopic lattice structure
% are encoded in the height representation
% and in the stiffness parameter $K$.
% Given these, everything can be understood in terms of
% the universal behavior of massless Gaussian fields.

In particular, when the $q$-state zero-temperature Potts antiferromagnet
on a two-dimensional lattice ${\cal L}$ admits a height representation,
one expects that $q = q_c({\cal L})$.
This prediction is confirmed in all heretofore-studied cases:
3-state square-lattice \cite{Nijs_82,Kolafa_84,Burton_Henley_97,Salas_98},
3-state Kagom\'e \cite{Huse_92,Kondev_96},
4-state triangular \cite{Moore_00},
and 4-state on the line graph (= covering lattice) of the square lattice
\cite{Kondev_95,Kondev_96}.

We now wish to observe that the height mapping employed for
the 3-state Potts antiferromagnet on the square lattice \cite{Salas_98}
carries over unchanged to any planar lattice in which
all the faces are quadrilaterals.
One therefore expects that $q_c = 3$
for every (periodic) plane quadrangulation.

\begin{figure}
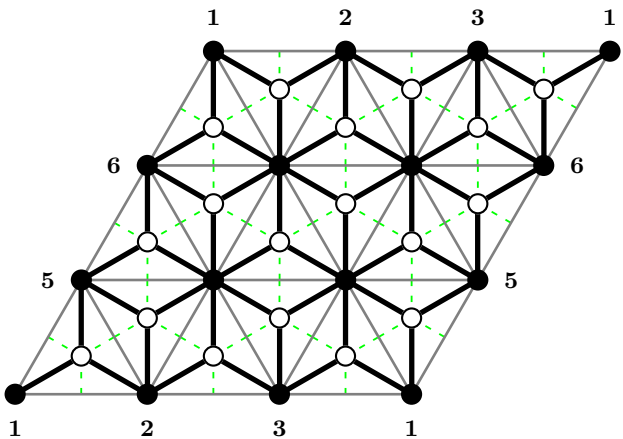

\centering
\psset{xunit=50pt}
\psset{yunit=50pt}
\psset{labelsep=10pt}
\pspicture(-0.1,-0.4)(4.6,3.0)
%%%%\psframe  (-0.5,-0.5)(5.0,3.1)

%% OLD VERSION OF PICTURE
%%
%\multirput{0}(0,0)(0.5,0.866025){3}{% 
%   \multirput{0}(0,0)(1,0){3}{%
%     \psline[linewidth=1pt,linestyle=dashed,linecolor=green](0.5,0.288675)(0.5,0)
%     \psline[linewidth=1pt,linestyle=dashed,linecolor=green](0.5,0.288675)(0.25,0.433013)
%     \psline[linewidth=1pt,linestyle=dashed,linecolor=green](0.5,0.288675)(0.75,0.433013)
%     \psline[linewidth=1pt,linestyle=dashed,linecolor=green](1,0.57735)(1,0.866025)
%     \psline[linewidth=1pt,linestyle=dashed,linecolor=green](1,0.57735)(0.75,0.433013)
%     \psline[linewidth=1pt,linestyle=dashed,linecolor=green](1,0.57735)(1.25,0.433013)
%     \psline[linewidth=1pt,linestyle=dotted,linecolor=red](0,0)(0.5,0.866025)(1,0)(0,0)
%     \psline[linewidth=2pt,linecolor=black](0,0)(0.5,0.288675)(0.5,0.866025) 
%     \psline[linewidth=2pt,linecolor=black](0.5,0.288675)(1,0)
%   }
%   \psline[linewidth=1pt,linestyle=dotted,linecolor=red](3,0)(3.5,0.866025)
%}
%\multirput{0}(0,0)(0.5,0.866025){3}{% 
%   \multirput{0}(0.5,0.866025)(1,0){3}{%
%     \psline[linewidth=2pt,linecolor=black](0,0)(0.5,-0.288675)(0.5,-0.866025) 
%     \psline[linewidth=2pt,linecolor=black](0.5,-0.288675)(1,0)
%   }
%}
%\psline[linewidth=1pt,linestyle=dotted,linecolor=red](1.5,2.59808)(4.5,2.5980)

\multirput{0}(0,0)(0.5,0.866025){3}{%
   \multirput{0}(0,0)(1,0){3}{%
      \psline[linewidth=0.7pt,linestyle=dashed,dash=2pt 3pt,linecolor=green](0.5,0.288675)(0.5,0)
      \psline[linewidth=0.7pt,linestyle=dashed,dash=2pt 3pt,linecolor=green](0.5,0.288675)(0.25,0.433013)
      \psline[linewidth=0.7pt,linestyle=dashed,dash=2pt 3pt,linecolor=green](0.5,0.288675)(0.75,0.433013)
      \psline[linewidth=0.7pt,linestyle=dashed,dash=2pt 3pt,linecolor=green](1,0.57735)(1,0.866025)
      \psline[linewidth=0.7pt,linestyle=dashed,dash=2pt 3pt,linecolor=green](1,0.57735)(0.75,0.433013)
      \psline[linewidth=0.7pt,linestyle=dashed,dash=2pt 3pt,linecolor=green](1,0.57735)(1.25,0.433013)
     \psline[linewidth=1pt,linecolor=gray](0,0)(0.5,0.866025)(1,0)(0,0)
     \psline[linewidth=2pt,linecolor=black](0,0)(0.5,0.288675)(0.5,0.866025)
     \psline[linewidth=2pt,linecolor=black](0.5,0.288675)(1,0)
   }
   \psline[linewidth=1pt,linecolor=gray](3,0)(3.5,0.866025)
}
\multirput{0}(0,0)(0.5,0.866025){3}{%
   \multirput{0}(0.5,0.866025)(1,0){3}{%
     
\psline[linewidth=2pt,linecolor=black](0,0)(0.5,-0.288675)(0.5,-0.866025)
     \psline[linewidth=2pt,linecolor=black](0.5,-0.288675)(1,0)
   }
}
\psline[linewidth=1pt,linecolor=gray](1.5,2.59808)(4.5,2.5980)

\multirput{0}(0,0)(0.5,0.866025){4}{% 
   \multirput{0}(0,0)(1,0){4}{%
      \pscircle*(0,0){4pt}
   }
}
\multirput{0}(0,0)(0.5,0.866025){3}{% 
   \multirput{0}(0,0)(1,0){3}{%
      \pscircle*[linecolor=white](0.5,0.288675){4pt}
      \pscircle(0.5,0.288675){4pt}
      \pscircle*[linecolor=white](1,0.57735){4pt}
      \pscircle(1,0.57735){4pt}
   }
}

\uput[270](0,0){$\bm{1}$}
\uput[270](1,0){$\bm{2}$}
\uput[270](2,0){$\bm{3}$}
\uput[270](3,0){$\bm{1}$}

\uput[90](1.5,2.59808){$\bm{1}$}
\uput[90](2.5,2.59808){$\bm{2}$}
\uput[90](3.5,2.59808){$\bm{3}$}
\uput[90](4.5,2.59808){$\bm{1}$}

\uput[180](0.5,0.866025){$\bm{5}$}
\uput[180](1.0,1.73205){$\bm{6}$}

\uput[0](3.5,0.866025){$\bm{5}$}
\uput[0](4.0,1.73205){$\bm{6}$}
\endpspicture
\vspace*{-2mm}
\caption{
   A diced lattice of size $3\times 3$ with periodic boundary conditions
   (edges depicted with thick black lines).
   The full circles show the sites of degree 6, which form a triangular lattice
   %% (edges depicted with thin dotted red lines).
   (edges depicted with thin gray lines).
   The open circles show the sites of degree 3, which form a hexagonal lattice
   (edges depicted with thin dashed green lines)
   that is the dual of the triangular lattice.
   Periodic boundary conditions are implemented by identifying
   border sites with the same label.
}
\label{fig1}
\end{figure}

The diced lattice (Fig.~\ref{fig1})
is a periodic tiling of the plane
by rhombi having $60^\circ$ and $120^\circ$ interior angles;
in particular, it is a plane quadrangulation
in which all vertices have degree 3 or 6.
The diced lattice is the dual of the Kagom\'e lattice,
%% (Fig.~\ref{fig2}),
which is in turn the medial graph of the triangular and hexagonal lattices.

%%%%%%%%%%%%%%%%%%%%%%%%%%%%%%%%%%%%%%%%%%%%%%%%%%%%%%%%%%%%%%%%%%%%%%%%%%%%%%%%
%%% FIGURE 1 -- DICED LATTICE

%%%%%%%%%%%%%%%%%%%%%%%%%%%%%%%%%%%%%%%%%%%%%%%%%%%%%%%%%%%%%%%%%%%%%%%%%%%%%%%%

In this Letter we give a rigorous proof that the 3-state diced-lattice
Potts antiferromagnet has a phase transition at {\em nonzero}\/ temperature.
This shows that, contrary to theoretical expectations,
$q_c({\rm diced}) > 3$.
It provides, moreover, the first example of a {\em bipartite}\/
%% (= loose-packed)
two-dimensional lattice in which $q_c > 3$ (but see below).

% In this Letter we provide compelling Monte Carlo evidence 
% for a phase transition at {\em nonzero}\/ temperature
% in the 3-state Potts antiferromagnet on the diced lattice.
% In the conventional Potts variable $v = e^J - 1$,
% for which $v \ge 0$ corresponds to the ferromagnetic region
% and $-1 \le v \le 0$ corresponds to the antiferromagnetic region
% (with $v=-1$ being the zero-temperature antiferromagnet),
% we find that the phase transition occurs at
% $v_c = -0.8???? \pm 0.????$.
% {\bf What can we say about the critical exponents???}

Furthermore, this model provides a simple concrete example
of entropically-driven long-range order,
in which coexistence between regions of different types of
order on one sublattice is disfavored because it reduces
the freedom of choice of spins on the other sublattice.
Though this idea is intuitively appealing, it is difficult
to determine, in any specific case, whether the mechanism
is strong enough to produce long-range order.  Here we are
able to resolve this question rigorously by a conceptually
simple though numerically delicate Peierls argument.

The existence (though not the nature) of a phase transition
in this model
is not, however, totally unexpected.
A decade ago, Jensen {\em et al.}\/ \cite{Jensen_97}
computed low-temperature expansions for the 3-state and 4-state
Potts models on the Kagom\'e lattice
and found, among other things, indications of
singularities in the unphysical region
at $v = -3.486 \pm 0.003$ and $v = -3.38 \pm 0.06$, respectively
(here $v = e^J - 1$ where $J$ is the nearest-neighbor coupling
 and we take $\beta=1$).
%% For the 3-state model the critical exponents at this transition
%% were indistinguishable from those of a 3-state Potts ferromagnet,
%% while for the 4-state model no exponent estimates could be obtained.
%% 
Shortly thereafter, Feldmann {\em et al.}\/ \cite{Feldmann_98}
used the duality $v \mapsto q/v$ of $q$-state
Potts models on planar lattices to deduce predictions
for the singularities of the 3-state and 4-state
Potts models on the diced lattice:
%% in particular, the singularities corresponding to the
%% unphysical Kagom\'e singularities occur at
$v = -0.8607 \pm 0.0008$ and $v = -1.18 \pm 0.02$, respectively.
The latter occurs in the unphysical region at $v < -1$,
suggesting that the 4-state diced-lattice antiferromagnet
lies in a disordered phase at all temperatures, including zero temperature.
The former, by contrast, lies within the physical antiferromagnetic regime
at $J = -1.971 \pm 0.006$.
%% (here $v = e^{-\beta} - 1$ for the antiferromagnet).
If these predictions are correct, we have
%% the 3-state diced-lattice antiferromagnet
%% has a finite-temperature phase transition,
%% with exponents that are equal or nearly equal to those
%% of the 3-state ferromagnet.
$3 < q_c({\rm diced}) < 4$;
crude linear interpolation suggests $q_c({\rm diced}) \approx 3.4$.
%% (Surprisingly, the papers \cite{Jensen_97,Feldmann_98} received
%%  no follow-up, probably because the theoretical importance of this
%%  example was not recognized.)

% This prediction of a finite-temperature phase transition is curious,
% however, because the 3-state Potts antiferromagnet at zero temperature
% on any plane quadrangulation can be mapped onto a height model
% by exactly the same mapping that is used on the square lattice
% \cite{Salas_98}.
% Experience with height representations tells us that,
% whenever such a representation exists,
% the corresponding spin model is almost always
% (but not always) critical \cite{height_rep_exceptions}.
% One therefore expects that the 3-state Potts antiferromagnet
% on any plane quadrangulation --- and in particular on the diced lattice ---
% has its critical point at zero temperature ($v=-1$);
% in particular, one expects $q_c({\cal L}) = 3$ for all such lattices.
% The foregoing prediction of a finite-temperature antiferromagnetic transition
% suggests that this expectation is false for the diced lattice,
% and in particular that $3 < q_c({\rm diced}) < 4$.
% (Crude linear interpolation between the predicted 3-state and 4-state
%  critical points suggests that $q_c({\rm diced}) \approx 3.4$.)
One might, however, worry that the errors in the series extrapolation
are radically larger than estimated
and that the diced-lattice singularity lies not at $v \approx -0.86$
but instead at the theoretically expected $v = -1$
(which is not, after all, so far away).
It is thus important to obtain independent evidence on the location
of the phase transition (if any)
in the 3-state diced-lattice Potts antiferromagnet.
We do this in two steps:
a mathematically rigorous proof of a phase transition at nonzero temperature,
which furthermore elucidates the entropically-driven nature
of the corresponding long-range order;
and Monte Carlo simulations
%% using the Wang--Swendsen--Koteck\'y (WSK) cluster algorithm \cite{WSK}.
to locate the phase transition and investigate its properties.

\paragraph{Proof of phase transition.}
We shall prove that, at all sufficiently low temperatures,
there is antiferromagnetic long-range order
in which the spins on the triangular sublattice
take preferentially one value and the spins on the hexagonal
sublattice take more-or-less randomly the other two values.
To implement the heuristic idea
%% is that the cost of coexistence between regions of
%% unequal spins on the triangular sublattice is principally entropic,
%% i.e.\ it reduces the freedom of choice of spin on the hexagonal sublattice
%% on the boundary between such regions.
%% To evaluate this cost, we first integrate out
of entropically-driven order,
we evaluate the cost of coexistence between regions of unequal spins
on the triangular sublattice by integrating out
the spins on the hexagonal sublattice,
yielding a $q$-state Potts model on the triangular lattice
with 3-body interactions on the triangles, namely,
Boltzmann weights $(w_1,w_2,w_3) = (q+3v+3v^2+v^3, q+3v+v^2, q+3v)$
according as the triangle has 1, 2 or 3 distinct spin values \cite{Kotecky_85}.
We then apply a Peierls argument to this general triangular-lattice model
and prove that there exists ferromagnetic long-range\ order
in an open region of $(w_2/w_1, w_3/w_1)$-space
that includes the point $(w_2/w_1, w_3/w_1) = (1/2,0)$
corresponding to $q=3$ at zero temperature ($v=-1$).

Choose a large box $\Lambda$ and fix all boundary spins in the same state.
Given a spin configuration on the triangular lattice,
draw Peierls contours on the dual hexagonal lattice in the usual way,
i.e., separating unequal spin values on the triangular lattice.
We thus obtain a spanning subgraph of the hexagonal lattice
in which the vertices have degree 0, 2 or 3
according as the corresponding triangle has 1, 2 or 3 distinct spin values.

Consider first the case $w_3=0$
(this covers the zero-temperature 3-state diced-lattice model, i.e., $q=3$ and $v=-1$).
Then the Peierls contours have no vertices of degree 3,
so they are disjoint unions of self-avoiding polygons (SAPs) on the hexagonal lattice.
Each contour edge gets a weight $w_2/w_1$,
and each contour gets an additional weight $q-1$
to count the possible values for the spin change modulo $q$ when crossing the contour.
If the probability of having at least one contour is less than $(q-1)/q$,
then the spin at the origin has probability greater than $1/q$ of being in the same state
as the boundary condition, hence there is long-range order.
This occurs whenever
\begin{equation}
   \sum_{n=6}^\infty  q_n^{(1)} \, (w_2/w_1)^n  \;<\;  1/q  \;,
 \label{eq.peierls_condition}
\end{equation}
where $q_n^{(1)}$ is the number of $n$-step hexagonal-lattice SAPs
surrounding the origin of the triangular lattice,
or equivalently the first area-weighted moment for $n$-step hexagonal-lattice SAPs
modulo translation.
To bound this sum, we use the exact values of $q_n^{(1)}$ for $6 \le n \le 140$
\cite{Jensen_06}
and the bound $q_n^{(1)} \le (n^2/36) \, 1.868832^{n-2}$ for even $n \ge 142$
\cite{SAP_bound}.
For $q = 3$ we deduce long-range order whenever $w_2/w_1 < 0.503417$,
which barely includes the desired value $w_2/w_1 = 1/2$
\cite{SAP_bound_remark}.

The case $w_3 > 0$ is more complicated because of the presence
of degree-3 vertices,
but it can be shown \cite{KSS_in_prep} %% \cite{Kotecky-Sokal}
that if the Peierls inequality holds at a given value of $w_2/w_1$
when $w_3 = 0$, then it will also hold at that same value of $w_2/w_1$
whenever $w_3/w_1$ is sufficiently small.
This proves that the 3-state diced-lattice antiferromagnet
has long-range order on the triangular sublattice
at all sufficiently low temperatures.
With a little more work  \cite{KSS_in_prep}, %% \cite{Kotecky-Sokal},
we can prove antiferromagnetic long-range order on the whole diced lattice.

\paragraph{Monte Carlo simulation.}

We simulated the diced-lattice Potts antiferromagnets for $q=2,3,4$
on $L \times L$ lattices ($3 \le L \le 768$)
with periodic boundary conditions,
using the Wang--Swendsen--Koteck\'y (WSK) cluster algorithm \cite{WSK}.
Since the diced lattice is bipartite,
the WSK algorithm is guaranteed to be ergodic
\cite{Burton_Henley_97,Ferreira-Sokal}
and there is reason to hope that critical slowing-down might be absent
(as for the square lattice \cite{Ferreira-Sokal})
or at least small.

We measured the energy $\scre$,
the sublattice magnetizations $\scrm_{\rm hex}$ and $\scrm_{\rm tri}$,
and the second-moment correlation length $\xi$
\cite{note_second-moment}.
%% (see \cite{Salas_98}).
%% {\bf I guess it's too complicated to explain how we defined $\xi$???}
We focussed attention on the Binder-type ratio
$R = \<\scrms^4\> / \<\scrms^2\>^2$
where $\scrms = \scrm_{\rm tri} - \scrm_{\rm hex}$,
which tends in the infinite-volume limit
to $(q+1)/(q-1)$ in a disordered phase
and to 1 in an ordered phase,
and is therefore diagnostic of a phase transition.
The ratio $\xi/L$ plays a similar role.
Finally, we studied $\< \scrms^2 \>$
in order %%% DELETE THESE WORDS IF NECESSARY TO SAVE SPACE
to estimate the leading magnetic critical exponent.

% We began by making a ``coarse'' set of runs
% covering a wide range of $v$ values, using modest-sized lattices
% and modest statistics.  If the plots of $R$ or $\xi/L$ versus $v$
% indicated a likely phase transition, we then made a ``fine'' set of runs
% covering a small neighborhood of the estimated critical point,
% using larger lattices and larger statistics.
% Finally, using the results from these latter runs,
% we made a ``super-fine'' set of runs extremely close to
% the estimated critical point,
% using as large lattices and statistics as we could manage,
% with the goal of obtaining precise quantitative estimates
% of the critical point $v_c$ and the critical exponents.
% The complete set of runs reported in this Letter
% used approximately 0.95 yr CPU time on a
% 1.86 GHz Intel Core 2 E6320 processor.

%% The runs for the Ising case ($q=2$)
The Ising ($q=2$) data
confirm the exact solution \cite{Syozi_72}.
%% ; this is useful as a test of correctness of our programs.
For $q=4$, we find a finite correlation length
uniformly down to zero temperature,
with $\xi(v) \uparrow \approx 1.85$ as $v \downarrow -1$.

%%%%%%%%%%%%%%%%%%%%%%%%%%%%%%%%%%%%%%%%%%%%%%%%%%%%%%%%%%%%%%%%%%%%%%%%%%%%%%
%
% FIGURE 2
%

\begin{figure}
%\vspace*{0cm} \hspace*{-0cm}
\begin{center}
\includegraphics[width=0.85\columnwidth]{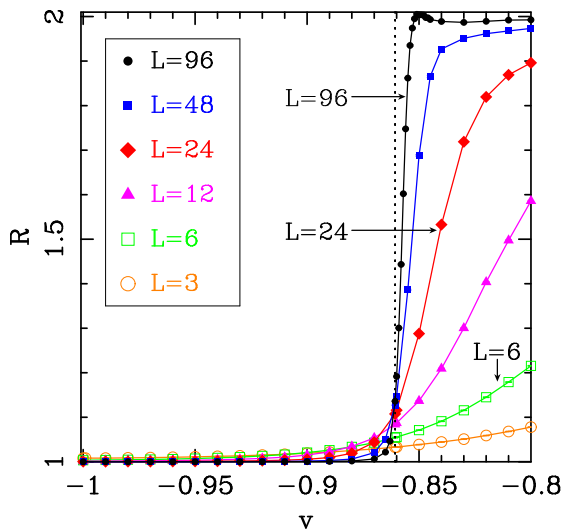}
%%\quad \vspace{6cm}  %% TEMPORARY UNTIL FILE IS THERE
\end{center}
\vspace*{-5mm}
\caption{
   Coarse plot for the Binder ratio $R$.
   Dotted vertical line marks the critical point predicted in
   \protect\cite{Jensen_97,Feldmann_98}.
   Curves are straight lines connecting points, meant only to guide the eye.
}
\label{fig_coarse_R}
\end{figure}
%%%%%%%%%%%%%%%%%%%%%%%%%%%%%%%%%%%%%%%%%%%%%%%%%%%%%%%%%%%%%%%%%%%%%%%%%%%%%%

%%%%%%%%%%%%%%%%%%%%%%%%%%%%%%%%%%%%%%%%%%%%%%%%%%%%%%%%%%%%%%%%%%%%%%%%%%%%%%
%
% FIGURE 3
%

\begin{figure}
%\vspace*{0cm} \hspace*{-0cm}
\begin{center}
\includegraphics[width=0.81\columnwidth]{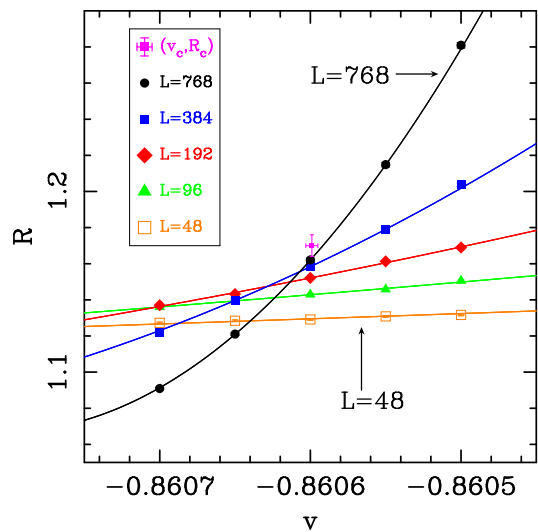}
%%\quad \vspace{6cm}  %% TEMPORARY UNTIL FILE IS THERE
\end{center}
\vspace*{-5mm}
\caption{
   %% Super-fine
   Fine plot for the Binder ratio $R$.
   %% Dotted vertical line marks the critical point predicted in
   %% \protect\cite{Jensen_97,Feldmann_98}.
   Curves are our fits to \reff{eq.R_Ansatz}.
   Symbol $\,\protect\fancyplus\,$ indicates estimates of $v_c$ and $R_c$.
}
\label{fig_superfine_R}
\end{figure}
%%%%%%%%%%%%%%%%%%%%%%%%%%%%%%%%%%%%%%%%%%%%%%%%%%%%%%%%%%%%%%%%%%%%%%%%%%%%%%

%%%%%%%%%%%%%%%%%%%%%%%%%%%%%%%%%%%%%%%%%%%%%%%%%%%%%%%%%%%%%%%%%%%%%%%%%%%%%%
%
% FIGURE 4
%

\begin{figure}
%\vspace*{0cm} \hspace*{-0cm}
\begin{center}
\includegraphics[width=0.81\columnwidth]{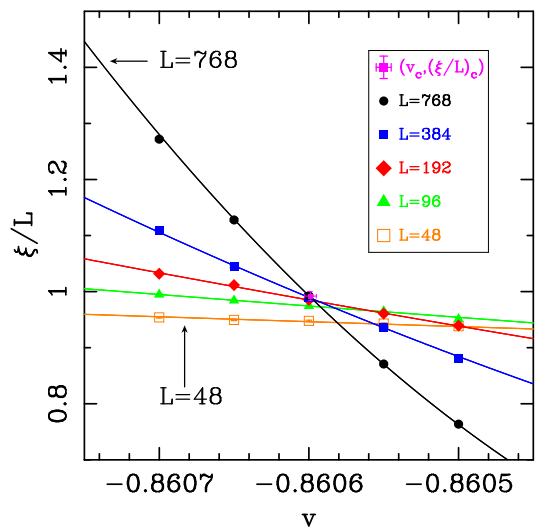}
%%\quad \vspace{6cm}  %% TEMPORARY UNTIL FILE IS THERE
\end{center}
\vspace*{-5mm}
\caption{
   %% Super-fine
   Fine plot for $\xi/L$.
   Curves are our fits to \reff{eq.R_Ansatz}.
   Symbol $\,\protect\fancyplus\,$ indicates estimates of
   $v_c$ and $(\xi/L)_c$.
}
\label{fig_superfine_xioverL}
\end{figure}
%%%%%%%%%%%%%%%%%%%%%%%%%%%%%%%%%%%%%%%%%%%%%%%%%%%%%%%%%%%%%%%%%%%%%%%%%%%%%%

For $q=3$,
%% the ``coarse'' plot of $R$
a plot of $R$
%% {\bf (or $\xi/L$??? or both???)}
versus $v$ for lattice sizes $3 \le L \le 96$
is shown in Fig.~\ref{fig_coarse_R},
and shows a clear order-disorder transition at $v_c \approx -0.86$.
%% The ``super-fine'' plots of $R$ and $\xi/L$,
Finer plots of $R$ and $\xi/L$ near the transition, for
%% lattice sizes   %% COMMENTED OUT TO SAVE SPACE FOR EQUATION (2) ON ONE LINE
                   %% DOESN'T WORK NOW, BUT MAY WORK IN PUBLISHED VERSION
$48 \le L \le 768$,
are shown in Figs.~\ref{fig_superfine_R} and \ref{fig_superfine_xioverL}.
We fit the data to Ans\"atze obtained from
\begin{eqnarray}
   & &
   \scro \;=\;  \scro_c \,+\, a_1 (v-v_c) L^{1/\nu}
                \,+\, a_2 (v-v_c)^2 L^{2/\nu}
        \qquad \nonumber \\
   & & \qquad\qquad
     \:+\; b_1 L^{-\omega_1} \,+\,
     %%  b_2 L^{-\omega_2} \,+\,
     \ldots
 \label{eq.R_Ansatz}
\end{eqnarray}
by omitting various subsets of terms,
and we systematically varied $L_{\rm min}$
(the smallest $L$ value included in the fit).
We also made analogous fits for $\< \scrms^2 \> / L^{\gamma/\nu}$.
Comparing all these fits,
we estimate the critical point $v_c = -0.860599 \pm 0.000004$,
the critical exponents $\nu = 0.81 \pm 0.02$
and $\gamma/\nu = 1.737 \pm 0.004$,
and the universal amplitude ratios $R_c = 1.170 \pm 0.007$
and $(\xi/L)_c = 0.995 \pm 0.007$
(68\% subjective confidence intervals, including both statistical error
 and estimated systematic error due to unincluded corrections to scaling).
These exponents are in excellent agreement with the values for the
3-state Potts ferromagnet:
$\nu = 5/6 \approx 0.833$, $\gamma/\nu = 26/15 \approx 1.733$
and $R_c = 1.1711 \pm 0.0005$ \cite{R_c_ferro}.
This confirms our expectation that the 3-body-interacting
triangular-lattice ferromagnet, obtained by integrating out
the hexagonal sublattice, lies in the universality class
of the 3-state Potts ferromagnet.
%% {\bf Maybe we should also give fits for $v_c$, $R_c$ and $(\xi/L)_c$
%%   with the exponents $\nu = 5/6$ and $\gamma/\nu = 26/15$ fixed.}
% {\bf How to estimate systematic error on $\gamma/\nu$ and $R_c$???
%    Maybe do fits with $w_c$ and $y_t$ \emph{imposed} at a variety of
%    plausible values, and see how much $\gamma/\nu$ and $R_c$ move???}
%%% A finite-size-scaling plot of $R$
%%% {\bf (or $\xi/L$??? or both???)}
%%% using these parameters is shown in Figure~\ref{fig_R_FSS}.
%% A ``coarse'' plot of $\< \scrms^2 \> / L^{\gamma_{\rm stagg}/\nu}$
%% using the estimated value of $\gamma_{\rm stagg}/\nu$
%% is shown in Figure~\ref{fig_chistagg_coarse}.

For the triangular-sublattice spontaneous magnetization $M_0$,
defined by $M_0^2 = \lim_{L\to\infty} \< \scrm_{\rm tri}^2 \> / V_{\rm tri}^2$,
we find $M_0 = 0.936395 \pm 0.000006$ at $v=-1$,
which is not far from the Peierls bound
$M_0 \ge 0.90497$ \cite{note_Peierls_bound}
and the heuristic estimates $M_0 \approx 125/128 \approx 0.97656$
and $M_0 \approx 21/22 \approx 0.95455$
\cite{KSS_in_prep}.  %%  \cite{Kotecky-Sokal,Salas-Sokal_MC}.
% {\bf The first heuristic estimate says that one out of 32 TRI sites
%    will be surrounded by six HEX sites of all the same color;
%    and half of these TRI sites will then be of a different color
%    from the boundary condition.  So the probability of a TRI site
%    following the boundary condition is $p = 63/64$,
%    which implies $M_0 = (qp-1)/(q-1) = 125/128$.
%    Why does this underestimate the ``bad'' sites by more than
%    a factor of 2???}
% {\bf The second heuristic estimate is based on fixing to state 1
%    all the TRI sites that are at distance 2 from a given TRI site,
%    and computing the Gibbs measure inside.}

%% Details of these simulations will be reported elsewhere
Details of the simulations will be reported later
\cite{KSS_in_prep}.  %% \cite{Salas-Sokal_MC}.

\paragraph{Final remarks.}
Here is another construction that produces bipartite planar lattices
(not, however, plane quadrangulations) with $q_c > 3$
and indeed with $q_c$ arbitrarily large.
Let $\scrl$ be any lattice, and let $\scrl_2$ be the lattice obtained from $\scrl$
by subdividing each edge into two edges in series.
Then the Potts series law $(v_1,v_2) \mapsto v_1 v_2/(q+v_1+v_2)$ \cite{Sokal_bcc2005}
implies that the Potts model on $\scrl_2$ has a phase transition
whenever $v^2/(q+2v) = v_{\rm crit,ferro,\scrl}(q)$.
In particular, if $v_{\rm crit,ferro,\scrl}(q) < 1/(q-2)$,
then there is a solution $v \in (-1,0)$, so that $q_c(\scrl_2) > q$.
For instance, the triangular lattice $\scrt$ has a ferromagnetic critical point
when $v^3 + 3v^2 - q = 0$, from which we conclude that $q_c(\scrt_2) \approx 3.117689$.
Furthermore, Wierman \cite{Wierman_02} has constructed
periodic plane triangulations $T(k)$
[obtained by subdividing triangles in the triangular lattice]
whose bond percolation thresholds tend to zero as $k\to\infty$;
and the Potts series-parallel laws show, more generally,
that for each $q \ge 1$
one has $\lim_{k \to\infty} v_{{\rm crit,ferro},T(k)}(q) = 0$ \cite{note_FKG}.
It follows that $\lim_{k \to\infty} q_c(T(k)_2) = +\infty$.

\begin{acknowledgments}
%% We thank Chris Henley for correspondence concerning height representations,
%% and Neal Madras and Gordon Slade for correspondence concerning
%% self-avoiding polygons.
We thank Chris Henley, Neal Madras and Gordon Slade
for very helpful correspondence.
We especially thank Cris Moore for discussions some years ago concerning
the Peierls argument for Potts antiferromagnets;  he independently suggested
to consider the diced lattice.
%% We also thank an anonymous referee for helpful suggestions.
This work was supported in part by NSF grant PHY--0424082,
Spanish MEC grants MTM2005--08618 and FIS2004-03767,
and Czech grants GA{\v C}R 201/06/1323 and MSM~0021620845.
We thank the Isaac Newton Institute at the University of Cambridge,
where this work was completed.
\end{acknowledgments}

%%%%%%%%%%%%%%%%%%%%%%%%%%%%%%%%%%%%%%%%%%%%%%%%%%%%%%%%%%%%%%%%%%%%%%


\begin{thebibliography}{99}
\bibitem{SAP_bound_remark}  Most of the sum of this series comes from the
crude estimate of the tail.
If we use instead for $n \ge 142$ the estimated actual behavior
$q_n^{(1)} \approx (1/4\pi) (2+\sqrt{2})^{n/2} n^{-1}$ \cite{Jensen_06},
the condition \reff{eq.peierls_condition} holds for
$w_2/w_1 \ltapprox 0.541180$.
\bibitem{note_second-moment}  $\xi$ is an antiferromagnetic analogue
of \cite[eq.~(3.11)]{Salas_98}.
\bibitem{note_Peierls_bound}  This bound is based on the
estimated actual behavior
$q_n^{(1)} \approx (1/4\pi) (2+\sqrt{2})^{n/2} n^{-1}$ \cite{Jensen_06}
for $n \ge 142$.
\bibitem{note_FKG}  The passage from Wierman's $G(k)$ to $T(k)$ uses the
FKG inequality, which needs $q \ge 1$.
\end{thebibliography}
\end{document}